\documentclass[a4paper,fleqn]{amsart}
\usepackage{booktabs}
\usepackage{lscape}
\usepackage{algorithmicx}
\usepackage{algpseudocode}

\title[Iterated Local Search for $Pm|\ |\sum w_jT_j$]{%
  Iterated local search and very
  large neighborhoods for the parallel-machines total tardiness
  problem} 
\author{F.~Della~Croce, T.~Garaix, A.~Grosso} 
\date{}

\newcommand{\set}[1]{\left\{#1\right\}}

\newcommand{\Brack}[1]{\left<\text{#1}\right>}
\newcommand{\proc}[1]{\textsc{#1}}

\newenvironment{framedalgo}[1]{%
  \endgraf
  \begin{quote}
      #1
    \begin{algorithmic}[1]
}
{\end{algorithmic}
\end{quote}
}

\begin{document}

\begin{abstract}
  We present computational results with a heuristic algorithm for the 
  parallel machines total weighted tardiness problem. The algorithm 
  combines generalized pairwise interchange neighborhoods, dynasearch 
  optimization and a new machine-based neighborhood whose size is 
  non-polynomial in the number of machines. The computational results 
  significantly improve over the current state of the art 
  for this problem.
\end{abstract}

\maketitle

\section{Introduction}
\label{sec:intro}
In the $Pm|\ |\sum_j w_jT_j$ problem a set of jobs $N=\set{1,2,\dots,n}$ are
given, with their processing times $p_j$, weights $w_j$ and due dates
$d_j$.  The jobs are to be processed in a schedule $S$ on a set
$M=\set{1,2,\dots,m}$ of identical parallel machines so that their
  completion times $C_j$ minimize the objective function
\[T(S)=\sum_{j=1}^nw_jT_j=\sum_{j=1}^nw_j\max\set{C_j-d_j,0}.\]
For $m=1$ the problem reduces to the single-machine total tardiness problem, 
that is well studied and solved in both the exact and heuristic frameworks 
--- we refer to~\cite{Bigr08,Pan07,Rodr08,Rodr08b} 
and~\cite{Con02,Ergu06,Gro04} for recent developments.\par

The literature seems to be fairly limited for the problem with parallel
machines; the most recent references are~\cite{Paol07, Bilg04, Koul97, 
Rodr08, Rodr08b} to the authors'~knowledge.\par

Iterated Local Search (ILS, see~\cite{Lou02} for a survey) is a local
search framework that can be seen as a tradeoff between the naive
multistart and complex metaheuristics.  In multistart a local search
driven optimization starts several times (often a huge number of
times) from randomly generated initial solutions, in order to achieve
a wide exploration of the solutions set.  In metaheuristics a number
of sophisticated devices (genetic crossover, short or long-term
memory, etc) are employed in order to escape poor local optima. In ILS
the search is simply restarted from a slightly perturbed version of
the best-known solution. With this type of restart, the starting point
of each local search is not completely random, and the perturbation
--- called ``kick'' --- aims at projecting the search ``not too far''
from previously explored local optima, without completely loosing
their partially optimized structure.\par
Very Large Neighborhood Search (VLNS) denotes local search methods
that define and explore complex neighborhoods for combinatorial
optimization problems; such neighborhoods are characterized by having
an exponential number of neighbor solutions --- with respect to the
problem size --- but can be explored in polynomial time by means of
exact or heuristic procedures (see~\cite{Ahu02}). \par
ILS is often successfully coupled with VLNS, hence moving the
complexity of the search from the overall algorithm to the
neighborhood exploration. We refer to~\cite{Con02, Ergu06} for a successful
application of such a VLNS technique (called {\em dynasearch\/}) to
the $1|\ |\sum w_jT_j$ problem.\smallskip\par

Rodrigues et al.~\cite{Rodr08} proposed a simple and quite effective
ILS algorithm for the $Pm|\ |\sum_jw_jT_j$ problem, using a local search 
based on pairwise interchange operators. That algorithm was tested on a batch of $100$ instances with $n=40,50$ and 
$m=2,4$ derived from a subset of the $ 1 || \sum w_j T_j$ problem instances  
available in the OR-library\footnote{\tt http://people.brunel.ac.uk/\~{}mastjjb/jeb/info.html}.
Notice that, 
on that batch, the algorithm was able to detect all but one optimal solutions.

This paper aims at defining an improved ILS algorithm for $Pm|\ |\sum_jw_jT_j$ 
by incorporating VLNS techniques. Particularly, we introduce a dynasearch 
optimization on each machine in the shop and a new ``Very Large'' 
neighborhood whose size is non-polynomial 
{\em in the number of machines.\/}\par

We illustrate the basic building blocks of the algorithm and present
computational experiments for assessing their effectiveness in
Section~\ref{sec:GPI}. The complete algorithm is described in
Section~\ref{sec:the-algo}, where also the computational results are discussed.
 The proposed ILS algorithm outperforms the
ILS of~\cite{Rodr08} on instances with a number of jobs $n$ ranging
from $40$ to $300$, and a number of machines $m$ ranging from $2$ to $20$.
The advantage of the new algorithm grows on instances with large $m$
thanks to the new neighborhood.

\section{The basic neighborhoods}
\label{sec:GPI}
\subsection{Generalized pairwise interchanges}
The well-known GPI operators work on a sequence of jobs 
$\sigma$ producing a new sequence $\sigma^\prime$ 
Let $\sigma=\alpha i\pi j\omega$, with jobs $i$ and $j$ in position 
$k$ and $l$ respectively.  The most common GPI operators are
\begin{enumerate}
\item {\em Swap\/} $\alpha i\pi j\omega\to
  \alpha j\pi i\omega$ ($\pi$ may be empty);
\item {\em Forward insertion\/} $\alpha i\pi j\omega\to
  \alpha\pi ji\omega$;
\item {\em Backward insertion\/} $\alpha i\pi j\omega\to
  \alpha ji\pi\omega$.
\item {\em Twist\/} $\alpha i\pi j\omega\to\alpha j\bar\pi i\omega$ 
  with $\bar\pi=\pi$~reversed.
\end{enumerate}
The implementation of such operators is straightforward in single-machine 
sequencing problems with regular cost functions, since the machine is never 
idle and the sequence $\sigma$ {\em is\/} the schedule.
The so-called GPI dynasearch neighborhood for single-machine
sequencing problems combines possibly many {\em independent\/} moves of types 
(1)--(4);
two moves are said to be {\em independent\/} if the pairs of positions
$(k,l)$ and $(p,q)$ on which they act are non-overlapping, i.e
$\max\set{k,l}<\min\set{p,q}$. In a single machine environment with an
additive objective function the contributions of independent moves
combine additively, and the best set of independent moves can be
worked out by dynamic programming (see~\cite{Con02} for details).  A
GPI dynasearch neighborhood exploration for an $n$ jobs sequence requires
$\mathcal O(n^2)$ time with its best implementation (see~\cite{Ergu06}).

In parallel-machines environments, GPI operators can be applied provided 
that a sequence $\sigma$ can be converted to a schedule.
Rodrigues et al.~\cite{Rodr08} proposed a simple yet quite effective
ILS algorithm for the $Pm|\ |\sum_jw_jT_j$ problem. The algorithm
applies GPI operators --- limited to (1)--(3) in their implementation --- 
on a sequence of jobs; the schedule on
parallel machines associated with this sequence $\sigma=(j_1,j_2,\dots,j_n)$ is
computed from scratch by means of the most natural {\em dispatching
  rule:\/} assign the next job in the sequence to the earliest available
machine. The neighborhood exploration is performed with a
first-improve strategy, and frequent restarts are applied (one kick
every five complete descents).  \smallskip\par

Whereas the basic GPI neighborhood can be easily adapted to the
parallel machines environment, this is not the case for the GPI
dynasearch neighborhood: since the job starting times are determined
by applying the dispatching rule, the contribution of independent
moves is no longer purely additive. Rodrigues et al.~\cite{Rodr08} do
not provide a different notion of independent moves, neither it is easy
to see an obvious one.

\subsection{Integrating GPIs on parallel machines and Dynasearch}
A possible drawback of the basic GPI neighborhood is that, in a
parallel environment, the working sequence on each single machine is
poorly optimized, since the machine-sequencing criterion is
extremely crude. We then investigated the opportunity of adding a
single-machine optimization phase through the use of a dynasearch
neighborhood. We tested the following algorithms, called~1 and~2
respectively.
\begin{description}
\item[Algorithm~A1] The GPI iterated local search of~\cite{Rodr08} 
	(kindly provided by the authors).
\item[Algorithm~A2] The same algorithm, where after building a schedule
  by the dispatching rule, each single machine is optimized by a full
  descent using the GPI dynasearch neighborhood where moves (1)--(4) are 
  used.
\end{description}
We considered a batch of 125 randomly instances with $n=100$, $m=4$;
we refer to Section~\ref{sec:comp-res} for details on the generation
scheme.  Two parameters $R,T$ determine the practical difficulty of
the instances.  We recorded the performances of the algorithms in
terms of time spent for reaching the best solution and number of local
search descents performed. For both algorithms we allowed one hour of
CPU time.  Table~\ref{tab:gpi-gpidyn} points to a comparison of the
computational costs of the two algorithms in terms of CPU time and
number of descents, detailing them by $(R,T)$ pairs.  Out of the 125
instances in the batch, Algorithm~A2 delivered better solutions in 37
cases, and worse solutions in 27 (columns labeled ``\#Bests'').  The
higher number of better solutions comes at the cost of higher CPU
times to be spent in the search.  The number of descents required to
reach the best solution is always consistently less for Algorithm~A2 than for
Algorithm~A1, but Algorithm~A2 --- quite expectedly --- exhibits in most
cases higher CPU times, since every solution undergoes a full
dynasearch descent on each machine.  Anyway, in the details of the
tests we were able to observe that on 18 instances of the batch,
Algorithm~A2 finds a better solution {\em and\/} requires less CPU time
to reach the optimum; this behaviour comes out with dramatic evidence
on some classes of instances like $R=0.6, T=0.6$, and the classes with
$R=1.0$.  This test suggests that an effort for keeping highly
optimized sequences on the machines can be worth, if a clever search
strategy can be developed in order to limit the growth of the CPU time

\subsection{A parallel-machines neighborhood}
A simple notion of independent moves arises if, instead of using the
dispatch rule on a sequence, one works directly on the
jobs-to-machines assignment. Assume a schedule is given. 
We consider, for a given pair of machines
$m_1$, $m_2$, the respective job working sequences $\sigma_1$,
$\sigma_2$ and the following moves:
\begin{enumerate}
\item[(a)] extract a job $j$ from $\sigma_1$ and insert it into $\sigma_2$;
\item[(b)] extract a job $j$ from $\sigma_2$ and insert it into $\sigma_1$;
\item[(c)] extract jobs $i\in\sigma_1$, $j\in\sigma_2$ and insert $i$ in 
  $\sigma_2-\{j\}$, and $j$ in $\sigma_1-\{i\}$.
\end{enumerate}
The insertion position for a job $j$ in a sequence $\sigma$ is chosen
in linear time, so that $\sigma=\alpha j\omega$ and $T(\alpha
j\omega)$ is as small as possible.
We note that if such moves are executed on disjoint pairs of machines
$(m_1,m_2)$, $(m_3,m_4)$, \dots, they are {\em independent\/} moves,
in the sense that their contributions to the schedule's cost combine
additively.
Hence a neighborhood whose size is non-polynomial in the number of
machines can be defined as follows, for a given schedule.
\begin{enumerate}
\item For each pair of machines $(m_1,m_2)$, compute the maximum
  decrease in tardiness $\Delta_{m_1,m_2}$ that can be obtained by
  applying moves (a), (b) and~(c) to $\sigma_1$, $\sigma_2$, for all
  $j\in\set{\sigma_1}\cup\set{\sigma_2}$.
\item Build a weighted improvement graph $G(M,E)$ where 
  \[E=\big\{\{m_1,m_2\}\colon \Delta_{m_1,m_2}>0\big\}.\]
\item A neighbor schedule is generated by taking a matching on $G$ and
  executing the moves associated to the matching edges.
\end{enumerate}
Note that, for each pair of machines:
\begin{itemize}
\item no more than $\mathcal O(n)$ jobs have to be considered for moves (a) 
  and (b);
\item no more than $\mathcal O(n^2)$ pairs $i,j$ have to be considered for 
  move (c);
\item evaluating each move of type (a), (b), (c) requires $\mathcal O(n)$ operations 
 for each job (or job pair) --- note that the schedule defines the machine to which the 
 job is assigned.
\end{itemize}
Hence evaluating all the possible moves (a), (b), (c) requires $\mathcal O(n^3)$ operations.
The best neighbor schedule can be computed by taking a maximum
weighted matching in $G$; the whole process for building $G$ and
selecting the matching can be implemented with running time
\[\mathcal O(n^3)+\mathcal O(m^3)\approx\mathcal O(n^3)\qquad\text{(as $m<n$ in non trivial instances)}.\]

Simple testing showed that the solutions provided by the GPI ILS technique 
of~\cite{Rodr08} are often not locally optimal with respect to the parallel 
machines neighborhood. Particularly, applying one search of the parallel 
machine neighborhood on the best solution provided by Algorithm~A1 after 
a 1-hour run, we found 11 improvements (over 125 solutions) for the instances 
in the $n=100, m=10$ batch; 
the number of improvement rises to 
96 for the $n=300, m=20$ batch.


The exploration of the parallel neighborhood is
usually fast (less than 0.6 seconds on the $n=300$ instances).  
Hence, we keep the parallel machines neighborhood as a cheap
and useful tool whose impact becomes more and more important on instances 
with many machines.

\section{Combining different neighborhoods}
\label{sec:the-algo}
\subsection{Neighborhoods and refinements}
In view of the experimental observations reported in the previous section, 
a careful combination of GPI moves, dynasearch optimization and the parallel 
machines neighborhood can be the key instrument for handling larger 
instances of the $Pm|\ |\sum_jw_jT_j$ problem.
We used three neighborhoods called $N_1$, $N_2$, $N_3$.\par
\begin{itemize}
\item Neighborhood $N_1$ is the GPI neighborhood of Rodrigues et 
  al.~\cite{Rodr08}. Note that in~\cite{Rodr08} a somewhat sophisticated 
  rule is used for breaking ties in choosing the best neighbor. 
  We avoided it in favor of a random tie-breaking rule.
\item Neighborhood $N_2$ is the parallel-machines neighborhood 
  described in Section~\ref{sec:GPI}.
\item Neighborhood $N_3$ is a GPI neighborhood where every neighbor schedule 
  generated by a GPI operator is improved by a dynasearch descent applied on 
  each machine schedule.
\end{itemize}

In order to reduce the computational effort, we applied the following
refinements to $N1$ and $N_3$. 
Following~\cite{Rodr08}, the neighborhood 
search proceed by first-improve, hence the first improving neighbor is 
adopted as new solution. We observed that often profitable moves happen 
between jobs that appear in relatively ``close'' positions. Hence the GPI 
operators are applied in ``stages''; each stage applies the GPI operators 
between jobs in positions 
\[\text{$i$ and $i+\gamma\mod n$}, \qquad
\text{for $i=1,\dots,n$, $\gamma$ fixed}.\]
In exploring $N_1,N_3$, at each successive stage $\gamma$ is set to 
$1,2,\dots,n-1$. The improving neighbor is often found at early stages.

Accordingly with \cite{Rodr08}, a full exploration of a $N_1$ neighborhood 
is accomplished in $\mathcal O(n^3\log m)$ operations. An exploration 
of $N_2$ requires $\mathcal O(n^3)$ operations (see Section \ref{sec:GPI}).
Exploring $N_3$ takes up to $\mathcal O(n^4)$ operations.

\subsection{The algorithm}
We now describe the complete proposed algorithm.  It uses all three
neighborhoods $N_1$, $N_2$, $N_3$.  Among these neighborhoods, the
local search phase for $N_1$, $N_2$ is exploited to a full descent,
while for $N_3$ it is limited to a single neighborhood exploration
because of the higher computational cost of such procedure. This is
denoted in the pseudocode by the operations called
$\proc{FullDescent}$ and $\proc{Search}$.\par
The search starts from the usual EDD (dispatched) sequence, which is
often accepted in literature as a reasonable quick-and-dirty starting
point.  A full descent of $N_1$ performs the first optimization in the
main loop, then $N_2$ and one exploration of $N_3$ are invoked
iteratively one after the other, as long as the iteration is
profitable. The $\proc{Kick}$ phase ---
i.e. perturbation of the best-known solution --- consists of a limited
number of random swaps, removal and insertions performed among 
different machines.\par
The algorithm uses a time-limit as stopping criterion, since we
considered it the most simple and tunable one.

\begin{framedalgo}{\textbf{Algorithm~A3}}
  \State{Set $S^\ast:=S:=\Brack{The EDD dispatched sequence}$;}
  \State{Set $\proc{IterCount}:=0$;}
  \Repeat
    \State{Improve $S$ by executing dynasearch on each machine;}
    \State{$S^1:=\proc{FullDescent}(S,N_1)$;}
    \Repeat
      \State{$S^2:=\proc{FullDescent}(S^1,N_2)$;}
      \State{$S^3:=\proc{Search}(S^2,N_3)$;}
      \If{$ T(S^3)<T(S^2)$}
        \State{Set $S^1:=S^3$;}
      \EndIf
    \Until{$\Brack{$N_3$ failed improving $S^2$}$;}
    \If{$T(S^3)\geq T(S^\ast)$}
      \State{Set $\proc{IterCount}:=\proc{IterCount}+1$;}
    \Else
      \State{Set $S^\ast:=S^3$;}
      \State{Set $\proc{IterCount}:=0$;}
    \EndIf
    \If{$\Brack{\text{$N_2$ and $N_3$ failed to improve $S^1$}}$}
      \If{$\proc{IterCount}>\proc{MaxNoImprove}$}
        \State{$S:=\proc{Kick}(S^\ast)$;}
        \State{$\proc{IterCount}:=0$;}
      \Else
        \State{$S:=\proc{Kick}(S^3)$;}
      \EndIf
      \Else
        \State{$S:=S^3$;}
    \EndIf
  \Until{$\Brack{Time-limit exceeded}$}
\end{framedalgo}

Following \cite{Rodr08}, a kick is executed each $\proc{MaxNoImprove}$ 
non-improving iterations, with $\proc{MaxNoImprove}=5$. Also, as a further 
refinement, the number of stages in the exploration on $N_3$ was fixed to 
a maximum of $\gamma_{\max}=5$; note that a higher value of $\gamma_{\max}$ 
causes more time to be spent in exploring $N_3$ and, correspondingly, a lower number of 
kicks executed in the allowed time limit. The values $\proc{MaxNoImprove}=5$, 
$\gamma_{\max}=5$ gave the best results in some preliminary tests --- only a modest amount 
of testing was needed to identify this value, without need for extensive calibration.
The value $\gamma_{\max}=5$ actually lowers the time spent for 
exploring $N_3$ to $\mathcal O(\gamma_{\max}n^3)$.

\subsection{Evaluation of the algorithm}\label{sec:comp-res}
We tested the hybrid ILS algorithm on batches of random instances
adapted from the well established literature on tardiness problems in
single-machine environments.  The single-machine instances are
characterized by uniformly distributed random data with 
processing times $p_i$ and weights $w_i$ from $[1,100]$, and due dates 
from a uniform distribution whose bounds are determined by two parameters 
$R$, $T$ called {\em due date range\/} and {\em  tardiness factor\/} 
--- see for example References~\cite{CraPVW98,PVW85}. Specifically, the $d_i$ values 
are randomly drawn from 
\[\textstyle [(1-T-R/2)\sum_{i=1}^np_i, (1-T+R/2)\sum_{i=1}^np_i].\]
For fixed $n,m$, we considered five instances
for each $(R,T)$ pair, with $R,T\in\set{0.2,0.4,0.6,0.8,1.0}$ --- 25
$(R,T)$ pairs, and 125 instances.  For $n=40,50,100$ we used times,
weights, and due dates from the $375$ single-machine OR-library
instances. For $n=300$ we used the instances from Tanaka et
al.~\cite{Tan09}. The number of machines $m$ was set to $2$, $4$
and~$10$, and pushed up to $20$ for the largest instances.  The due
dates are adapted to the parallel machines case by scaling the due
dates by $\tfrac1m$ (rounding down the obtained values).  

Tables~\ref{tab:final}--\ref{tab:final2} focus on the comparison of
Algorithm~A1 by Rodrigues et al.~\cite{Rodr08} and Algorithm~A3.
Accordingly
with~\cite{Rodr08} we allowed at most one hour CPU time to each test 
and report an aggregated comparison in Table~\ref{tab:final}
\footnote{per-instance results are available as a compressed file 
at\\ {\tt www.di.unito.it/\~{}grosso/solution.tar.7z}};
the results obtained within shorter CPU times are also presented in
Tables~\ref{tab:final1} and~\ref{tab:final2}.  The total number of
instances tested was $1625$.  For both A1 and A3 the results of a single run 
are reported; 
although the kick phase accounts for some nondeterminism in A3, we did not 
observe it delivering significantly different tardiness values in different 
runs, as far as the time limits reported in the tables are allowed.\par  
In Table~\ref{tab:final} we report information on the behavior of the
algorithms running with a 3600 seconds time limit.  
The ``dev'' columns report the percentage average and
maximum deviations of the objective function delivered by A1 and A3
with respect to the best found value --- the ``best'' value is defined
as the minimum between the tardiness value obtained by the two
algorithms within the allowed 1-hour run. 
The column 
$\#_{\rm best}$ counts the number of instances (out of the 125) where
Algorithm~A1 or Algorithm~A3 delivered a better
solution. $\text{CPU}_\text{best}$ reports the average time-to-best
for both algorithms and $N_{\text{desc}}$ the average number of
descent performed.  The performances of the two algorithms are
basically comparable for ``small'' instances (say for $n=40,50$); with
such limited problem sizes, both A1 and A3 often find an optimal
solution.

Tables~\ref{tab:final1} and~\ref{tab:final2} compare the behavior of
the two algorithms for different values of the time limits
enforced.  The $n=300$ cases are not
reported for time limits $\leq30$ secs (Table~\ref{tab:final1})
because in several instances such time limits where not enough to
perform a full execution of A3 --- this is due to the heavier
computational requirements of the dynasearch component of A3.  Aside
from this limitation, A3 is seen to strongly outperform A1 in terms of
solution quality.

A3 becomes apparently the best option for large $n$ ($n=100,300$), 
and especially for instances with a large number of machines ($m=10,20$). 
On the latter instances the key factors for the success of A3 are
the ability to exploit the parallel-machines ``very large'' neighborhood, and
the powerful dynasearch neighborhood for optimizing each machine sequence.

\begin{table}[p]
  \[\begin{array}{llrrrcrrr}\toprule
    \multicolumn2{c}{}&\multicolumn3{c}{\text{Algorithm~2}}&
    &\multicolumn3{c}{\text{Algorithm~1}}\\
    R&T&{\rm CPU}_{\rm avg}&N{\text{desc}_{\rm avg}}&\text{\#Bests}&
    &{\rm CPU}_{\rm avg}&N{\text{desc}_{\rm avg}}&\text{\#Bests}\\ 
    \cmidrule{3-5}\cmidrule{7-9}
    0.2&0.2	&21.17	&3.60	&0	&&1.36	&4.40	&0	\\ 
    0.2&0.4	&27.65	&3.20	&0	&&7.10	&27.00	&0	\\ 
    0.2&0.6	&1164.99	&147.80	&1	&&301.41	&1266.40	&0	\\ 
    0.2&0.8	&2246.95	&252.00	&1	&&540.10	&3793.40	&0	\\ 
    0.2&1.0	&1690.30	&205.20	&2	&&723.50	&6689.40	&0	\\ 
    0.4&0.2	&121.81	&21.00	&0	&&5.43	&22.60	&0	\\ 
    0.4&0.4	&46.43	&4.80	&0	&&25.07	&99.20	&0	\\ 
    0.4&0.6	&752.06	&90.80	&2	&&1167.06	&5924.60	&1	\\ 
    0.4&0.8	&1662.31	&178.60	&1	&&1057.19	&8114.60	&4	\\ 
    0.4&1.0	&1370.86	&165.20	&2	&&976.78	&8721.60	&3	\\ 
    0.6&0.2	&4.00	&0.00	&0	&&0.07	&0.00	&0	\\ 
    0.6&0.4	&143.21	&19.00	&0	&&163.62	&614.40	&0	\\ 
    0.6&0.6	&414.91	&45.00	&2	&&1472.98	&8029.20	&0	\\ 
    0.6&0.8	&2075.77	&223.00	&3	&&2085.35	&14456.20	&2	\\ 
    0.6&1.0	&901.48	&91.40	&3	&&2178.29	&18166.40	&2	\\ 
    0.8&0.2	&4.20	&0.00	&0	&&0.07	&0.00	&0	\\ 
    0.8&0.4	&546.14	&84.40	&0	&&210.41	&890.80	&1	\\ 
    0.8&0.6	&1808.82	&196.40	&2	&&1781.18	&10169.60	&2	\\ 
    0.8&0.8	&2378.09	&244.40	&4	&&1562.70	&10504.00	&1	\\ 
    0.8&1.0	&3004.82	&339.80	&3	&&1149.06	&9383.20	&1	\\ 
    1.0&0.2	&4.05	&0.00	&0	&&0.07	&0.00	&0	\\ 
    1.0&0.4	&591.58	&84.00	&0	&&420.00	&2359.80	&0	\\ 
    1.0&0.6	&1459.21	&147.00	&3	&&1855.63	&10910.00	&2	\\ 
    1.0&0.8	&1015.41	&107.20	&5	&&2073.00	&13795.40	&0	\\ 
    1.0&1.0	&1879.91	&210.40	&3	&&2338.92	&17907.40	&2	\\ 
    \bottomrule
  \end{array}\]
  \caption{Basic GPI local search (Algorithm~1) and GPI+dynasearch (Algorithm~2).
   Comparison for $n=100$, $m=4$.}
  \label{tab:gpi-gpidyn}
\end{table}

\begin{table}[p]
\centerline{\begin{tabular}{rrrrrrrrrrrrrrrr}\toprule
 $n$ & $m$ && \multicolumn{4}{c}{dev$_\%$} &&\multicolumn{2}{c}{\# best}  
  && \multicolumn{2}{c}{CPU$_{\text{best}}$(sec)} && \multicolumn{2}{c}{$N_{\text{desc}}$}\\
\cmidrule{4-7}\cmidrule{9-10}\cmidrule{12-13}\cmidrule{15-16}
    &   && A3$_{\text{avg}}$&A1$_{\text{avg}}$&A3$_{\text{max}}$&A1$_{\text{max}}$          && A3 & A1 && A3      & A1     && A3    & A1 \\ \midrule
 40 & 2 &&    0.0&0.0&0.0&0.0     && 1 & 0 && 0.0 & 0.2 && 464 & 4819 \\
 40 & 4 &&    0.0&0.0&0.0&5.5     && 2 & 0 && 0.9 & 0.1 && 2174 & 8072 \\
 40 & 10 &&   0.0&0.0&0.0&1.5     && 8 & 0 && 0.3 & 0.2 && 34780 & 64940 \\
 50 & 2 &&    0.0&0.0&0.0&0.0     && 0 & 0 && 0.1 & 0.1 && 486 & 4414 \\
 50 & 4 &&    0.0&0.0&0.0&0.6     && 4 & 1 && 2.5 & 1.5 && 3077 & 9885 \\
 50 & 10 &&   0.0&0.1&0.0&8.1     && 13 & 0 && 0.2 & 29.6 && 75898 & 111079 \\
 100 & 2 &&   0.0&0.0&0.0&0.0     && 14 & 0 && 52.3 & 35.3 && 5480 & 44630 \\
 100 & 4 &&   0.0&0.0&0.0&0.1     && 55 & 4 && 835.7 & 1030.2 && 8958 & 23856 \\
 100 & 10 &&  0.0&0.3&0.0&9.0     && 86 & 1 && 1525.6 & 1139.7 && 11113 & 11562 \\
 300 & 2 &&   0.0&0.0&0.1&0.3     && 72 & 5 && 2800.9 & 3308.3 && 183 & 1192 \\
 300 & 4 &&   0.0&0.2&0.0&11.3    && 89 & 3 && 3290.4 & 1148.8 && 279 & 628 \\
 300 & 10 &&  0.0&0.8&0.0&21.3    && 102 & 0 && 2819.9 & 3022.6 && 346 & 287 \\
 300 & 20 &&  0.0&0.8&0.0&28.3    && 104 & 1 && 1663.1 & 2149.4 && 336 & 207 \\
\bottomrule
  \end{tabular}}
  \caption{Overall results (1 hour time limit).}
  \label{tab:final}
\end{table}


\begin{table}[p]
\centering
\begin{tabular}{rr||rrrr|rrrr|rrrr|}
 $n$ & $m$ & \multicolumn{4}{c|}{dev$_\%$ 5 s.} & 
 \multicolumn{4}{|c|}{dev$_\%$ 10 s.} & \multicolumn{4}{|c|}{dev$_\%$ 30 s.}\\
 \multicolumn2{c||}{}&
 \multicolumn{2}{c}{avg} & \multicolumn{2}{c|}{max}&
 \multicolumn{2}{c}{avg} & \multicolumn{2}{c|}{max} &
 \multicolumn{2}{c}{avg} & \multicolumn{2}{c|}{max} \\ 
   &      &A3 &A1&A3&A1&A3 &A1&A3&A1&A3 &A1&A3&A1\\ \hline
 40& 2    &0.0&0.0&0.0&0.4   &0.0&0.0&0.0&0.0  &0.0&0.0&0.0&0.0 \\
 40& 4    &0.1&0.1&0.0&11.3  &0.1&0.1&0.0&6.1  &0.0&0.1&0.2&6.1 \\
 40& 10   &0.3&0.6&0.2&22.2  &0.2&0.5&0.1&22.2 &0.2&0.4&0.1&13.9\\
 50& 2    &0.0&0.0&0.0&0.3   &0.0&0.0&0.0&0.2  &0.0&0.0&0.0&0.0 \\
 50& 4    &0.1&0.1&0.0&3.1   &0.0&0.0&0.0&3.1  &0.0&0.0&0.0&1.8 \\
 50& 10   &0.2&0.7&0.0&35.0  &0.1&0.6&0.0&35.0 &0.1&0.4&0.0&20.3\\
 100& 2   &0.0&0.0&0.0&2.0   &0.0&0.0&0.0&0.2  &0.0&0.0&0.0&0.2 \\
 100& 4   &0.1&0.4&0.0&29.1  &0.1&0.4&0.0&29.1 &0.1&0.3&0.0&29.1\\
 100& 10  &0.9&2.0&0.1&45.3  &0.6&1.5&0.0&44.1 &0.4&1.2&0.0&36.4\\
\end{tabular}
\caption{Comparison for small time limits ($\leq30$ sec).}
\label{tab:final1}
\end{table}

\landscape
\begin{table}[p]
\[\begin{tabular}{rr||rrrr|rrrr|rrrr|rrrr|rrrr|}
 $n$ & $m$ &\multicolumn{4}{c|}{dev$_\%$ 60 s.} &  \multicolumn{4}{|c|}{dev$_\%$ 120 s.} &  \multicolumn{4}{|c|}{dev$_\%$ 180 s.} 
 & \multicolumn{4}{|c|}{240 s.} & \multicolumn{4}{|c|}{dev$_\%$ 300 s.}\\
                  &                    &                    \multicolumn{2}{c}{avg}        & \multicolumn{2}{c|}{max}&      \multicolumn{2}{c}{avg}        & \multicolumn{2}{c|}{max} & \multicolumn{2}{c}{avg}    & \multicolumn{2}{c|}{max} & \multicolumn{2}{c}{avg}    & \multicolumn{2}{c|}{max} & \multicolumn{2}{c}{avg}    & \multicolumn{2}{c|}{max}\\
                        &                                        &                     A3        &          A1                    & A3  & A1  & A3  & A1  & A3  & A1  & A3  & A1  & A3  & A1  & A3  & A1  & A3  & A1      & A3  & A1 & A3  & A1\\\hline
 40 & 2     & 0.0 & 0.0 & 0.0 & 0.0 & 0.0 & 0.0 & 0.0 & 0.0 & 0.0 & 0.0 & 0.0 & 0.0 & 0.0 & 0.0 & 0.0 & 0.0 & 0.0 & 0.0 & 0.0 & 0.0 \\
 40 & 4     & 0.0 & 0.0 & 0.0 & 5.5 & 0.0 & 0.0 & 0.0 & 5.5 & 0.0 & 0.0 & 0.0 & 5.5 & 0.0 & 0.0 & 0.0 & 5.5 & 0.0 & 0.0 & 0.0 & 5.5  \\
 40 & 10    & 0.1 & 0.3 & 3.2 & 11.1 & 0.1 & 0.2 & 2.8 & 5.6 & 0.0 & 0.2 & 2.8 & 5.6 & 0.0 & 0.2 & 2.8 & 5.6 & 0.0 & 0.2 & 0.0 & 5.0  \\
 50 & 2     & 0.0 & 0.0 & 0.0 & 0.0 & 0.0 & 0.0 & 0.0 & 0.0 & 0.0 & 0.0 & 0.0 & 0.0 & 0.0 & 0.0 & 0.0 & 0.0 & 0.0 & 0.0 & 0.0 & 0.0 \\
 50 & 4     & 0.0 & 0.0 & 0.0 & 1.8 & 0.0 & 0.0 & 0.0 & 1.8 & 0.0 & 0.0 & 0.0 & 0.6 & 0.0 & 0.0 & 0.0 & 0.6 & 0.0 & 0.0 & 0.0 & 0.6 \\
 50 & 10    & 0.1 & 0.3 & 3.3 & 16.3 & 0.0 & 0.2 & 1.8 & 16.3 & 0.0 & 0.2 & 1.8 & 16.3 & 0.0 & 0.2 & 1.4 & 16.3 & 0.0 & 0.2 & 1.0 & 16.3 \\
 100 & 2    & 0.0 & 0.0 & 0.1 & 0.1 & 0.0 & 0.0 & 0.0 & 0.1 & 0.0 & 0.0 & 0.0 & 0.1 & 0.0 & 0.0 & 0.0 & 0.1 & 0.0 & 0.0 & 0.0 & 0.1 \\
 100 & 4    & 0.0 & 0.3 & 2.1 & 29.1 & 0.0 & 0.2 & 2.1 & 13.3 & 0.0 & 0.2 & 0.2 & 13.3 & 0.0 & 0.1 & 0.2 & 3.5 & 0.0 & 0.1 & 0.2 & 3.5 \\
 100 & 10   & 0.3 & 1.0 & 6.7 & 29.7 & 0.2 & 0.9 & 4.2 & 29.7 & 0.1 & 0.6 & 3.5 & 9.6 & 0.1 & 0.6 & 3.5 & 9.0 & 0.1 & 0.6 & 2.6 & 9.0  \\
 300 & 2    & 0.1 & 0.1 & 1.7 & 1.8 & 0.1 & 0.1 & 0.9 & 0.9 & 0.0 & 0.0 & 0.7 & 0.9 & 0.0 & 0.0 & 0.5 & 0.9 & 0.0 & 0.0 & 0.4 & 0.9 \\
 300 & 4    & 0.5 & 1.3 & 16.4 & 76.2 & 0.3 & 1.3 & 16.4 & 76.2 & 0.3 & 1.2 & 16.4 & 76.2 & 0.3 & 0.7 & 13.6 & 18.1 & 0.3 & 0.6 & 13.6 & 17.9 \\
 300 & 10   & 0.7 & 3.2 & 17.3 & 72.6 & 0.6 & 2.1 & 17.2 & 41.9 & 0.6 & 1.6 & 17.2 & 37.5 & 0.5 & 1.5 & 15.8 & 37.5 & 0.4 & 1.5 & 15.8 & 37.5\\
 300 & 20   & 0.7 & 4.3 & 11.4 & 96.1 & 0.6 & 2.2 & 11.4 & 50.4 & 0.5 & 2.0 & 11.4 & 50.4 & 0.5 & 1.9 & 11.4 & 50.4 & 0.4 & 1.8 & 10.3 & 50.4\\
  \end{tabular}\]
  \caption{Comparison for large time limits}
  \label{tab:final2}
\end{table}

\endlandscape

\begin{thebibliography}{99}


\bibitem{Ahu02} Ahuja~R.~K., Ergun~\"O., Orlin~J.~B., Punnen~A.~P.,
A survey of very large-scale neighborhood search techniques,
{\em  Discrete Applied Mathematics\/} {\bf 123}, 75--102 (2002).

\bibitem{Paol07} Anghinolfi~D., Paolucci~M., Parallel machine
    total tardiness scheduling with a new hybrid metaheuristic approach,
{\em   Computers and Operations Research} {\bf 34}, 3471--3490 (2007).

\bibitem{Bilg04} Bilge~U., Kyra\c c~S., Kurtulan~F., Pekgun~M., A
    tabu search algorithm for parallel machine total tardiness
    problem, {\em Computers and Operations Research} {\bf 31}, 397--414 (2004)

\bibitem{Bigr08} Bigras~L.~Ph., Gamache~M., Gilles~S.  Time-Indexed
  Formulations and the Total Weighted Tardiness Problem, {\em
    INFORMS Journal on Computing\/} {\bf 20}, 133--142 (2008).


\bibitem{Con02} Congram~R., Potts~C.~N., van~de~Velde~S., An
  iterated dynasearch algorithm for the single machine total weighted
  tardiness problem, {\em INFORMS Journal on Computing,\/} {\bf 14}, 
  52--67 (2002).

\bibitem{CraPVW98} Crawuels~H.~A.~J., Potts~C.~N., Van~Wassenhove~L.~N.,
  Local search heuristics for the single machine total weighted tardiness 
  scheduling problem, {\em INFORMS Journal on Computing\/} {\bf 10}, 
  341--350 (1998).

\bibitem{Ergu06} Ergun~\"O., Orlin~J.~B., Fast neighborhood search for
    the single machine total weighted tardiness problem, {\em Operations
  Research Letters\/} {\bf 34}, 41--45 (2006).

\bibitem{Gro04} Grosso~A., Della~Croce~F., Tadei~R., 
  An enhanced dynasearch neighborhood for the single machine total tardiness 
  problems, {\em Operations Reasearch Letters,\/} {\bf 32}, 68--72 (2004).

\bibitem{Koul97} Koulamas~C., Decomposition and hybrid simulated
    annealing heuristics for the parallel-machine total tardiness
    problem, {\em Naval Research Logistics\/} {\bf 44}, 109--125 (1997).

\bibitem{Lou02} Lourenco~H.~R., St\"utzle~T., 
  Iterated local search, in {\em Handbook of MetaheuRistics,\/} 
  F.~Glover and G.~Kochenberger, ISORMS 57, pp.~321--253 (2002).

\bibitem{Pan07} Pan~Y., Shi~L., 
  On the equivalence of the max-min transportation lower bound and the 
  time-indexed lower bound for single machine scheduling problems, 
  {\em Mathematical Programming,\/} {\bf 100}, 543--559 (2007).

\bibitem{PVW85} Potts~C.~N., Van~Wassenhove~L.~N.,
  A branch and bound algorithm for the total weighted tardiness problem, 
  {\em Operations Research,\/} {\bf 33}, 363--377 (1985).

\bibitem{Rodr08} Rodrigues~R., Pessoa~A., Uchoa~E., Poggi de Arag\~ao M. 
  Heuristic algorithm for the parallel machine total weighted tardiness 
  scheduling problem, internal report 10/2008, Universidad Federal Fluminense 
  {\tt http://www.producao.uff.br/conteudo/rpep/volume82008/RelPesq\_V8\_2008\_10.pdf}

\bibitem{Rodr08b} Rodrigues~R., Pessoa~A., Uchoa~E., Poggi de Arag\~ao M.,
  Algorithms over Arc-time Indexed Formulations for Single and
    Parallel Machine Scheduling Problems, Optimization-Online, 
    {\tt http://www.optimization-online.org/DB\_HTML/2008/06/2022.html}.

\bibitem{Tan09} Tanaka S., Fujikuma S., Araki M. 
  An exact algorithm for single-machine scheduling without 
  machine idle time, {\em Journal of Scheduling,\/} {\bf 12}, 575--593 (2009).
\end{thebibliography}
\end{document}